\documentstyle[12pt]{article}
\def\hbeta{{{\beta} \over {2}}}
\begin{document}
\begin{titlepage}
\rightline{CERN-TH/95-100}
\rightline{arch-ive/9506208}
\vspace{0.5cm}
\noindent
{\Large{\bf {\tt WWGENPV} $-$ A Monte Carlo event generator \\
for four-fermion production \\
in {\boldmath $e^+ e^- \to W^+ W^- \to 4f$} } }
\vspace{1cm}
\bigskip\bigskip

\noindent
Guido MONTAGNA \\
{\it Istituto Nazionale di Fisica Nucleare, Sezione di Pavia } \\
{\it Via A.~Bassi n.~6 - 27100 PAVIA - ITALY} \\
\medskip\medskip

\noindent
Oreste NICROSINI~\footnote{\footnotesize
On leave from INFN, Sezione di Pavia, Italy} \\
{\it CERN, Theory Division} \\
{\it CH 1211 - Geneva 23 - SWITZERLAND} \\
\medskip\medskip

\noindent
Fulvio PICCININI  \\
{\it Istituto Nazionale di Fisica Nucleare, Sezione di Pavia } \\
{\it Via A.~Bassi n.~6 - 27100 PAVIA - ITALY} \\

\noindent
Program classification: 11.1 \\
\bigskip\bigskip

\noindent
{\small The Monte Carlo program {\tt WWGENPV}, designed for computing
distributions and generating events for
the four-fermion process $e^+ e^- \to W^+ W^- \to 4f$, is
described. It is based on the calculation
of the exact tree-level matrix element of the four-fermion reaction
and includes
initial-state radiation in the leading-log approximation within the
structure function approach. The program
 can be used in a two-fold way: as a Monte Carlo integrator for
weighted events, providing predictions for the total cross section,
the $W$ invariant-mass distribution, the radiative
energy and invariant-mass loss; as a true event generator of
unweighted events, useful for simulation purposes. }
\medskip\medskip

\begin{center}
Submitted to Computer Physics Communications
\end{center}

\vfil
\leftline{CERN-TH/95-100}
\leftline{April 1995}
\end{titlepage}

\vfil
\eject

\leftline{\Large{\bf PROGRAM SUMMARY}}
\vskip 15pt

\leftline{{\it Title of program:} {\tt WWGENPV}}
\vskip 8pt

\noindent
{\it Computer:} DEC VAX, HP/APOLLO, IBM/RS6000;
{\it Installation:} INFN, Sezione di Pavia, via A.~Bassi 6, 27100 Pavia, Italy
\vskip 8pt

\leftline{{\it Operating system:} VMS, UNIX}
\vskip 8pt

\leftline{{\it Programming language used:} FORTRAN 77}
\vskip 8pt

\noindent
\leftline{{\it Memory required to execute with typical data:} 500 KByte}
\noindent

\vskip 8pt

\leftline{{\it No. of bits in a word: } 32}
\vskip 8pt

\leftline{\it The code has not been vectorized}
\vskip 8pt

\noindent
{\it Subprograms used:} random number generator {\tt RANLUX}~[1];
routines from the CERN Program Library are also used.
\vskip 8pt

\leftline{{\it No. of lines in distributed program, including test data,
etc.:} 2135. }
\vskip 8pt

\leftline{\it Correspondence to:}
\leftline{MONTAGNA@PV.INFN.IT;}
\leftline{NICROSINI@VXCERN.CERN.CH, NICROSINI@PV.INFN.IT;}
\leftline{PICCININI@PV.INFN.IT}
\vskip 8pt


\vskip 8pt

\leftline{{\it Nature of physical problem} }
\noindent
The precise measurement of the $W$-boson mass $M_W$ constitutes a primary task
of the forthcoming experiments at the high energy electron--positron
collider LEP2 ($2 M_W \leq \sqrt{s} \leq 210$~GeV). This experimental goal
will be possible through the study of the four-fermion reaction
$e^+ e^- \to W^+ W^- \to 4f$.
A meaningful comparison between theory and
experiment requires an accurate description of
the fully exclusive process $e^+ e^- \to W^+ W^- \to 4f$, including the
main effects of radiative corrections, with the final goal of
providing predictions for the {\it realistic} distributions
measured by the experiments~[2].

\noindent
\vskip 8pt
\leftline{{\it Method of solution} }
\noindent
In order to efficiently perform the high-dimensional integration
 (due to the four-fermion final state) in the presence of realistic cuts,
 the total cross section, the energy and invariant-mass loss
and the $W$ invariant-mass distributions
 are computed by means of a Monte Carlo integration for weighted
events. For simulation purposes, the program can also be used as
an event generator that provides a sample of unweighted events,
 defined as the components of the four final-state fermions momenta, plus
the radiative variables of the incoming electron and positron, plus $\sqrt
{s}$,
stored into proper $n$-tuples. The importance-sampling technique~[3]
is employed to take care of the peaking behaviour of the integrand, both
in the integration and in the generation modes.

\vskip 8pt
\leftline{{\it Restrictions on the complexity of the problem} }
\noindent
Electroweak processes originating the same four-fermion final state
of the double-resonating diagrams
$e^+ e^- \to W^+ W^- \to 4f$ ({\it background} processes)
are not included in the present version of the program.
The treatment of the initial-state
radiation is limited to the gauge-invariant
subset of the leading-logarithmic corrections which are the
 most important contribution from the phenomenological point of view.
 Furthermore, the kinematics of the event in the presence of
the initial-state radiation is limited to the collinear approximation,
i.e. $p_T / p_L$ effects are neglected for the time being.

\vskip 8pt
\leftline{\it Typical running time}
\noindent
As integrator, the code needs about 5 h of HP 9000/735 for generating $10^8$
weighted events; the relative errors on the cross section and on the energy
and invariant-mass losses are about $2 \times 10^{-4}$ and $6 \times 10^{-4}$,
respectively. The generation of a sample of $10^4$ unweighted events requires
about 3 min on the same system.
\noindent

\vskip 8pt

\leftline{{\it Unusual features of the program:} none}
\noindent
\vskip 8pt

\leftline{{\it References} }
\noindent
\vskip 8pt \noindent
[1] F.~James, Comput. Phys. Commun. 79 (1994) 111.

\vskip 10pt \noindent
[2] G.~Montagna, O.~Nicrosini, G.~Passarino and F.~Piccinini,
Phys.~Lett. B348 (1995) 178, and references therein.

\vskip  10pt \noindent
[3] F.~James, Rep. Prog. Phys. 34 (1980) 1145.
\vfil
\eject

\leftline{\Large{\bf LONG WRITE -- UP}}
\vskip 15pt

\section{Introduction}
\vskip 10pt

The precise measurement of the $W$-boson mass $M_W$, and more
generally the study of its physical properties, constitutes a primary task
of the forthcoming experiments at the high energy electron-positron
collider LEP2 ($2 M_W \leq \sqrt{s} \leq 210$~GeV).

A large amount of theoretical work has been performed up to now, following
essentially two main streams.

On the one hand, one can find semi-analytical works, whose leitmotiv is
computing analytically as much phase-space integrations as
possible~\cite{sa}. Typically, these works give (QED-corrected) distributions
for the invariant masses of the $W^\pm$, integrating {\it a priori} all the
other independent phase-space variables, from which total cross sections can
be easily computed. After such an analytical effort, the semi-analytical
codes developed along this line are very fast, precise and efficient. They
are hence typically well suited for fitting purposes, their main drawback
being the lack of a full reconstruction of the true event phase-space
configuration.

On the other hand, one can find Monte Carlo codes. In principle they can
guarantee the full event reconstruction, since they handle the
four-momenta of the final-state
particles. Of course, the accuracy with which this goal is
achieved depends on the dynamics built-in in the code; in particular, the
exclusive control of all the features of the full four-fermion event,
including all the possible correlations between final-state particles, requires
the knowledge of the exact four-fermion matrix element, at least in the Born
approximation (for instance, no isotropic decays of the intermediate bosons
can be allowed).

Common to both the above-mentioned approaches, is the problem of treating QED
corrections, which, as is well known, are large and strongly dependent on the
experimental cuts applied. In particular, it happens that the energy lost in
the beam pipe (which at the leading-logarithmic level is well described by
initial-state radiation in the collinear approximation) represents an
important part of the systematic error on $M_W$. Therefore one needs also a
reliable model for, at least, initial-state radiation.

Some Monte Carlo codes are already available for the four-fermion processes
of interest at LEP2~\cite{mc}. Nevertheless, the critical comparison between
independent approaches to a problem has always been a useful and powerful tool
for understanding the theoretical accuracy of the results. The aim of this
work is to present and describe in some detail {\tt WWGENPV},
a Monte Carlo event
generator for $e^+ e^- \to W^+ W^- \to 4 f$, including initial-state QED
corrections in the structure functions formalism.

\section{Theoretical Background and General Features}
\vskip 10pt

The theoretical formulation implemented in the program
 has been described in some detail in~\cite{noiww}.
To calculate the cross section of
the $e^+ e^- \to W^+ W^- \to 4f$ process in the presence of the
initial-state radiation, the basic formula reads:
\begin{eqnarray}
\sigma (s) = \int d x_1 \, d x_2 \, D(x_1,s) D(x_2,s)
d [PS] {{d \sigma_0} \over {d [PS]}} .
\end{eqnarray}
In this equation, $d [PS]$ denotes the volume element in the
7-dimensional phase space connected to the independent variables of the
hard scattering reaction in the centre-of-mass
frame; ${{d \sigma_0} / {d [PS]}}$
is the exact differential kernel cross section in the Born
approximation, limited at present to the $W$-signal diagrams.
The inclusion of the
complete set of background diagrams is in progress. At any rate their effect
can be cut-off by proper selection criteria on the final-state
invariant masses.
Finally $D(x,Q^2)$ is the electron
structure function describing initial-state radiation~\cite{sf}.
For its explicit expression the following form is chosen:
\begin{eqnarray}
& &D(x,s) = \, { {\exp \left\{ \frac{1}{2}\beta \,\left( \frac{3}{4} -\gamma_E
\right) \right\} }
\over {\Gamma \left( 1 + \frac{1}{2} \beta \right) } } \,
\hbeta (1 - x)^{\hbeta - 1} - {{\beta} \over 4} (1+x) \nonumber \\
& & + {1 \over {32}} \beta^2 \left[ -4 (1+x) \ln(1-x) + 3 (1+x) \ln x -
4 {{\ln x} \over {1-x}} - 5 - x \right] ,
\end{eqnarray}

\noindent
with

\begin{eqnarray}
\beta = 2\,{\alpha \over \pi}\,(L-1) , \quad \quad \quad
\end{eqnarray}

\noindent
where $L= \ln \left( {s / {m^2}} \right)$ is the collinear logarithm,
$\gamma_E$ is the Euler constant and $\Gamma(z)$ is the gamma function.
The first exponentiated term of Gribov--Lipatov form
describes multiphoton soft radiation, the second and third ones hard
collinear bremsstrahlung. This choice of the structure function is a
consistent solution of the Renormalization Group evolution equation at
the leading-logarithmic level, and guarantees that, after an expansion
down to $\cal O(\alpha)$, the well-known factorized and hence universal
bremsstrahlung spectrum is exactly recovered.

With the choice of the independent variables adopted in~\cite{noiww},
the differential kernel cross section is proportional to the
 tree-level squared matrix element as follows:
\begin{eqnarray}
{{d \sigma_0} \over {d [PS]}} \propto {{\vert M \vert^2}\over
{\sqrt{(a \mu_3^4 + b \mu_3^2 + c)(a' \tau_2^2 + b' \tau_2 + c')}}} .
\label{eq:bornw}
\end{eqnarray}
Therefore, in the kernel cross section two kinds of ``singularities''
can be recognized, namely the one related to the vector boson propagators
(on $\mu_{1,2}^2$), the second arising from the Jacobian
appearing in~(\ref{eq:bornw}) (on $\mu_3^2, \tau_2$). To
efficiently perform the Monte Carlo integration in the presence of the
above peaking structure, the importance-sampling technique~\cite{james}
has been employed. For instance, for the $W$-boson propagator
singularities, an integration must be performed, which has the form
\begin{eqnarray}
I = \int_{x_m^2}^{x_M^2} dx^2\, {{f(x^2)}\over{(x^2-a^2)^2 + b^2}},
\end{eqnarray}
where $x^2 = \mu_i^2$, $a^2 = M_W^2/s$, $b^2 = \Gamma^2 M_W^2 / s^2$ and
the function $f(x^2)$ is regular.
Multiplying and dividing the integrand by the weight function
\begin{eqnarray}
p(x^2) = {{N^{-1}(x_m^2,x_M^2)}\over{(x^2-a^2)^2 + b^2}},
\end{eqnarray}
with
\begin{eqnarray}
N(x_m^2,x_M^2) &=& \int_{x_m^2}^{x_M^2} dx^2\, {1 \over{(x^2-a^2)^2 + b^2}}
\nonumber \\
&=& {1 \over b}\left[ \hbox{\rm Arctg}
\left( {{x_M^2 - a^2}\over b} \right)
- \hbox{\rm Arctg} \left( {{x_m^2 - a^2}\over b} \right) \right] ,
\end{eqnarray}
one obtains
\begin{eqnarray}
I = \int_{x_m^2}^{x_M^2} dx^2\, p(x^2) \, \left[ {{p^{-1}(x^2) f(x^2)}
\over{(x^2-a^2)^2 + b^2}} \right],
\end{eqnarray}
where the quantity in square brackets is no longer singular at
$x^2 = a^2$. The integral $I$ can be rewritten as
\begin{eqnarray}
I = \int_{0}^{1} dc\, {{p^{-1}(x^2(c)) f(x^2(c))}
\over{(x^2(c)-a^2)^2 + b^2}},
\end{eqnarray}
introducing the cumulative variable $c$ as
\begin{eqnarray}
c(x^2) = \int_{x_m^2}^{x^2} dz^2\, p(z^2) .
\end{eqnarray}
Therefore, by generating $c$ flat between $0$ and $1$ an accurate estimate
of $I$ can be obtained.

Concerning the singularities arising from the Jacobian factor,
 the integrations to be performed are of the following kind:
\begin{eqnarray}
I = \int_{y_m}^{y_M} dy\, {{g(y)}\over{\sqrt{(y-y_1)(y_2-y)}}}.
\end{eqnarray}
In this case the importance sampling can be applied
by introducing the weight function
\begin{eqnarray}
p(y) = N^{-1}(y_m,y_M) (y_2 - y)^{-{1\over 2}} (y - y_1)^{-{1\over 2}} ,
\end{eqnarray}
with
\begin{eqnarray}
N(y_m,y_M) &=& \int_{y_m}^{y_M} dy\, (y-y_m)^{-{1\over 2}}
(y_M-y)^{-{1\over 2}} \nonumber \\
&=& 2\left[ \hbox{\rm Arcsin}
\sqrt{ {{y_M - y_1}\over {y_2 - y_1}} }
- \hbox{\rm Arcsin} \sqrt{{{y_m - y_1}\over {y_2 - y_1}}} \right] .
\end{eqnarray}

When including initial-state QED corrections,
in order to perform the importance sampling on the generation of the radiative
variable $x$ in the soft region $x \to 1$,
the weight function is chosen as the all-order soft part
of the structure function, i.e.
\begin{eqnarray}
p(x) = N^{-1}(x_m, x_M) {{\beta }\over 2} (1 - x)^{{{\beta }\over 2} - 1} ,
\label{eq:isd}
\end{eqnarray}
with the  normalization factor $N(x_m, x_M)$ given by:
\begin{eqnarray}
N(x_m,x_M) &=& \int_{x_m}^{x_M} dy\,
{{\beta }\over 2} (1 - x)^{{{\beta }\over 2} - 1}  \nonumber \\
&=& (1 - x_m)^{{{\beta }\over 2}} - (1 - x_M)^{{{\beta }\over 2}}.
\end{eqnarray}
\noindent

With respect to the original formulation of QED corrections given in
\cite{noiww}, besides a different choice of the structure function,
the Coulomb singularity, which is important in the threshold
region, is included in the code according to the very recent
formulae given in~\cite{fmkc}. For details
and links with other approaches to the problem,
the reader is referred to the original literature.

\section{Program Structure}
\vskip 10pt

After the initialization of the Standard Model parameters and of
the electromagnetic quantities, the calculation within the
program proceeds as described in the following.
As a first step, the random number generator {\tt RANLUX}
is called to extract a vector of twelve components which are
converted from
{\tt REAL*4} to {\tt REAL*8} precision in order to be consistent with the
calculational environment, which is in {\tt REAL*8} precision.
The first two random numbers
 are used to generate the radiative variables $x_{1,2}$ according
to the distribution given by the soft exponentiated part of the
structure function; eight random numbers are employed to fully
reconstruct the kinematics of the four-fermion hard scattering
process in the centre-of-mass
frame: seven of them are needed to generate the independent variables
previously introduced for the kernel reaction, one to choose between the
two possible solutions of the non-linear constraint.
Given the independent variables, all the invariants related to the
$2 \to 4$ kernel reaction are computed.
The variables $\mu_{1,2,3}^2$ and
$\tau_{2}$ are, in particular, generated by means of the
importance-sampling procedure to take care of the
peaking behaviour of the integrand, as described
in the previous section. The other
variables ($z_{1,3}, \tau_1$) are generated according to
the uniform distribution in the physical range relative to an
extrapolated set-up. When initial-state radiation is switched on,
the four-momenta of the outgoing fermions are reconstructed
in the laboratory frame, taking into account the
Lorentz boost due to the emission of unbalanced electron and
positron radiation and given by the $\beta_L$ Lorentz factor
$\beta_L = (x_1 - x_2) / (x_1 + x_2)$. At this point,
 one further random number is
generated to perform a random
rotation around the beam axis to account for a
trivial overall azimuthal angle. The last random number is
required in the generation branch for hit-or-miss.
After that, a call to {\tt SUBROUTINE CUTUSER} is performed
 in order to implement the rejection algorithm on
the final-state particles. If the
event is accepted, the squared matrix element in Born
approximation is returned by a call to {\tt SUBROUTINE BORN}.
In conclusion, at the end of the loop over the events,
the averages and squared averages are cumulated to provide the
``weighted events'' Monte Carlo results for the required observables and
their numerical errors.

When the program runs as a generator of unweighted events,
after the hit-or-miss procedure,
the components of the four final-state fermions momenta, plus
the radiative variables of the incoming electron and positron, plus $\sqrt
{s}$,
are stored into a proper $n$-tuple. Each event stored in the $n$-tuple
has the structure $(x_1, x_2 ,E_b, \vec q_1, \vec q_2, \vec q_3,
\vec q_4 )$, where $E_b$ is the beam energy and $\vec q_i$ are the
three-momenta of the outgoing particles, assumed to be
ultrarelativistic. In this case, the cross section
for unweighted events is given as output.

\section{Input}
\vskip 10pt

Some input parameters and flags are introduced in the program
in order to allow the user to select the output of interest.
A short explanation of their meaning is addressed in the following and
is also printed when running the code interactively.

\noindent
{\bf \begin{verbatim}
OGEN(CHARACTER*1)
\end{verbatim}}
\noindent
It controls the use of the program as a Monte Carlo integrator
for weighted events ({\tt OGEN = I}) or as a Monte Carlo
event generator of unweighted events ({\tt OGEN = G}). If the first choice is
requested, the code returns in a LOGFILE
the value of the cross section for weighted
events, together with the average energy loss and invariant-mass loss,
defined as $\langle (2-x_1-x_2) \sqrt{s} /2 \rangle$ and
$\langle (1-x_1 x_2) \sqrt{s} /2 \rangle$, respectively. If the second
choice is requested, it prints
in a LOGFILE the value of the cross section for unweighted events and fills
an $n$-tuple, {\tt EXAMPLE.DAT},
containing the generated (unweighted) events. The
LOGFILE contains also the value of {\tt NBIAS/NMAX}. If it is $> 0$, the
generation procedure is biased and it is necessary to run the code again,
after having properly increased the value of
{\tt FMAX} (default = 1000).

\vskip 5pt

\noindent
{\bf \begin{verbatim}
RS(REAL*8)
\end{verbatim}}
\noindent
The centre-of-mass energy (in GeV).

\vskip 5pt

\noindent
{\bf \begin{verbatim}
NHITWMAX(INTEGER)
\end{verbatim}}
\noindent
Required by the integration branch. It gives the maximum number of
calls for the Monte Carlo loop.
\vskip 5pt

\noindent
{\bf \begin{verbatim}
NHITMAX(INTEGER)
\end{verbatim}}
\noindent
Required by the event-generation branch. It is the maximum number of
hits for the hit-or-miss procedure.
\vskip 5pt

\noindent
{\bf \begin{verbatim}
IQED(INTEGER)
\end{verbatim}}
\noindent
This flag allows the user to switch on/off the contribution of the
initial-state radiation. If {\tt IQED = 0} the distributions are
computed in lowest-order approximation, while for {\tt IQED = 1}
the initial-state QED corrections are included in the calculation.
This choice can be performed both for the integration and
for the generation mode.

\vskip 5pt

\noindent
{\bf \begin{verbatim}
ODIS(CHARACTER*1)
\end{verbatim}}
\noindent
Required by the integration branch.
It selects the kind of experimental distribution. For {\tt ODIS = T}
the program computes the total cross section
(in pb) of the process, eventually
including the cuts supplied by the user in {\tt SUBROUTINE CUTUSER} (see
below);
 for {\tt ODIS = W} (only possible with {\tt OGEN = I})
the $W$ invariant-mass distribution is returned (in pb/GeV).
In this case one can choose between the two opposite charged $W$-bosons,
 namely $W^-$ ({\tt IWCH = 1}) and $W^+$ ({\tt IWCH = 2}),
 and the invariant-mass of interest must be entered by the user in GeV.

\vskip 5pt

\noindent
{\bf \begin{verbatim}
OWIDTH(CHARACTER*1)
\end{verbatim}}
\noindent
It allows a different choice of the value of the $W$-width.
{\tt OWIDTH = Y} means that the tree-level Standard Model
formula for the $W$-width is used; {\tt OWIDTH = N} requires that
the $W$-width is supplied by the user in GeV. This freedom in the choice
of the $W$-width has been introduced so as to simplify the comparisons
of the results of our code with those of other programs.
\vskip 15pt

\noindent
{\bf \begin{verbatim}
NSCH(INTEGER)
\end{verbatim}}
\noindent
The value of
{\tt NSCH} allows the user to choose the calculational scheme for the
weak mixing angle and the gauge coupling. Three choices are available. If
{\tt NSCH=1}, the input parameters
used are $G_F, M_W, M_Z$ and the calculation is
performed at tree level. If {\tt NSCH = 2} or
{\tt 3}, the input parameters used are
$\alpha(Q^2), G_F, M_W$ or $\alpha(Q^2), G_F, M_Z$, respectively,
 and the calculation is performed using the QED coupling constant at a
proper scale $Q^2$, which is requested as further input.

\noindent
{\bf \begin{verbatim}
OCOUL(CHARACTER*1)
\end{verbatim}}
\noindent
This flag allows the user to switch on/off the contribution of the
Coulomb correction. If {\tt OCOUL = Y (N)} the distributions are
computed with (without) the above electromagnetic effect.

\vskip 5pt

\section{Test Run Output}
\vskip 10pt

The typical calculations that can be performed with {\tt WWGENPV}
are illustrated by the following examples of test run output.

\noindent
\begin{description}

\item[Sample 1] Calculation of total cross section, average energy loss and
average invariant-mass loss for $10^8$ weighted events, including
initial-state QED corrections.

\item[Sample 2] Calculation of total cross section for $10^4$ unweighted
events and filling the $n$-tuple {\tt EXAMPLE.DAT}, including
initial-state QED corrections.

\end{description}

\noindent
The output of sample~1 can be obtained with the following input:
\noindent

\begin{verbatim}
OGEN = 'I'
\end{verbatim}
\noindent
The integration branch is chosen.

\begin{verbatim}
RS = 176.D0
\end{verbatim}
\noindent
The centre-of-mass energy is fixed at 176~GeV.

\begin{verbatim}
NHITWMAX = 100000000
\end{verbatim}
\noindent
The maximum number of calls is fixed.

\begin{verbatim}
IQED = 1
\end{verbatim}
\noindent
Initial-state QED corrections are included.

\begin{verbatim}
ODIS = 'T'
\end{verbatim}
\noindent
The total cross section is computed (together with the corresponding average
energy and invariant-mass loss).

\begin{verbatim}
OWIDTH = 'Y'
\end{verbatim}
\noindent
The $W$-boson width is computed at the tree level in the Standard Model.

\begin{verbatim}
NSCH = 2
\end{verbatim}
\noindent
$\alpha(Q^2)$, $G_F$ and $M_W$ are used as input parameters.

\begin{verbatim}
1/ALPHAI = 128.07D0
\end{verbatim}
\noindent
The value of $1/\alpha(Q^2)$ is entered.

\begin{verbatim}
OCOUL = 'N'
\end{verbatim}
\noindent
The Coulomb correction is not included.

\vskip 12pt
\noindent
The output of sample~2 can be obtained with the following modifications:
\noindent

\begin{verbatim}
OGEN = 'G'
\end{verbatim}
\noindent
The generation branch is chosen.

\begin{verbatim}
NHITMAX = 10000
\end{verbatim}
\noindent
The number of generated events is fixed.

\noindent
Moreover the value of {\tt ODIS} is not required.

Both for the integration and generation branch, the output
returns the values of the ``Feynman rules'' {\tt GVE, GAE, GWF} and
{\tt GWWZ} linked to the vertices occurring in the diagrams
of $e^+ e^- \to W^+ W^- \to 4f$ as follows:
\begin{eqnarray}
& &Z_{f \bar f} \, \hbox{\rm vertex} = \gamma_{\mu}
\, ( \hbox{\tt GVE} + \hbox{\tt GAE} \, \gamma_5)
\nonumber \\
& &W_{u \bar d, l \bar {\nu_l}} \, \hbox{\rm vertex} =
\hbox{\tt GWF} \, \gamma_{\mu} \, (1 + \gamma_5) \nonumber \\
& &WWZ \, \hbox{\rm vertex} =
\hbox{\tt GWWZ} \times \,  \hbox{\rm Lorentz  structure} .
\nonumber
\end{eqnarray}

\vskip 15pt
\section{Program Implementation}
\vskip 10pt

In this section a short description of each routine implemented in the program
 is given. For random-number generation the routine {\tt RANLUX} by F.~James
is used and inserted in the code.
For more details about {\tt RANLUX} the reader is referred to
the proper literature~\cite{rlux}.

\vskip 20pt

\noindent
{\bf \begin{verbatim}
SUBROUTINE CUTUSER(EVENT,CUTFLAG)
REAL*8 CUTFLAG
DIMENSION EVENT(15)
\end{verbatim}}

\vskip 5pt\noindent
To be supplied by the user.
The subroutine allows the user to implement any kind of
experimental cut on the four-momenta of the final-state fermions.
Globally, the event is specified by 15 ingredients:
the radiative variables $x_{1,2}$ ({\tt EVENT(1), EVENT(2)}),
the beam energy ({\tt EVENT(3)}) and the components of the three-momenta
 of the four outgoing fermions ({\tt EVENT(I)}, with {$ I=4,15 $ }).
The energies are simply derived by the mass-shell condition for
ultrarelativistic fermions.
The reference frame is assumed to be with the $z$-axis along the direction of
the incoming positron.

The routine returns a flag ({\tt CUTFLAG = 0, 1}), which controls if the
event is accepted ({\tt CUTFLAG = 1}) or rejected ({\tt CUTFLAG = 0})
by the selection criteria.

\noindent
\vskip 20pt
{\bf \begin{verbatim}
SUBROUTINE BORN(S,X13,X14,X34,X35,X15,X16,X23,X24,
                X25,X26,X36,X45,X46,X56,SQM,IOVFL,IOEFL,
                IOES,JOEFL,JOES)
IMPLICIT REAL*8 (A-H,O-Z)
\end{verbatim}}
\vskip 5pt\noindent
It returns in {\tt SQM} the squared tree-level matrix element associated
to the $s$- and $t$-channels Feynman diagrams of the kernel
process $e^+ e^- \to W^+ W^- \to 4f$. The squared matrix element
 is computed numerically, squaring the amplitude obtained by means of
 {\tt SCHOONSCHIP}~\cite{schoon} in the framework of the helicity-amplitude
 formalism~\cite{hel} for massless final fermions.
If the effect of the Coulomb correction is
required, the value returned for {\tt SQM} is given by:
\begin{eqnarray}
\hbox{\tt SQM} =
\hbox{\tt SQM} \cdot \left\{1 + \, \alpha / v \, \delta_C \right\}  ,
\end{eqnarray}
where $\delta_C$ represents the Coulomb correction factor and $v$ the relative
 velocity of the two $W$-bosons (see~\cite{fmkc} for more details and
explicit formulae).

\noindent
\vskip 20pt
{\bf \begin{verbatim}
FUNCTION DOP(X)
IMPLICIT REAL*8 (A-H,O-Z)
COMMON/EM/RL,ETA,BETA,SDELTA
\end{verbatim}}
\vskip 5pt\noindent
The QED structure function, adopted to account for the radiation emitted
by the incoming leptons, is computed according to the
analytical expression reported above. It is given by the sum
of an exponentiated term of the Gribov--Lipatov form, associated to soft
multiphoton emission, with up to ${\cal O}(\alpha^2)$ contributions
describing hard bremsstrahlung in collinear approximation. Actually,
in order to perform importance sampling on the generation of the radiative
variable $x$ in the soft region $x \to 1$, the following formula is
implemented in {\tt FUNCTION DOP}:
\begin{eqnarray}
\overline{D} (x) \, = \, D(x) \, / p(x) ,
\end{eqnarray}
where $p(x)$ is the weight function given by~(\ref{eq:isd}).

\vskip 15pt

\leftline{\Large{\bf Acknowledgements}}
\vskip 10pt
The authors would like to thank the INFN, Sezione di Pavia,
 for having provided computer resources. They are also grateful
to Giorgio Fumagalli and Valerio Vercesi
for helpful assistance in numerical and graphical problems.
Useful discussions and exchanges of information
with D.~Bardin and D.~Charlton are gratefully acknowledged.

\vfil
\eject
\pagestyle{empty}

\leftline{\bf TEST RUN OUTPUT}
\vskip 12pt

\leftline{\bf Sample 1}

\begin{verbatim}

 SQRT(S) =  176.0  GEV
 M_Z =  91.1888  GEV
 G_Z =  2.4974  GEV
 M_W =  80.23  GEV
 G_W =  2.03367033063195  GEV
 STH2 =  0.2310309124510679

 GVE =  -1.409737267299689E-02
 GAE =  -0.185794027211796
 GWF =  0.230409927395451
 GWWZ =  -0.571479454308384

 RANLUX INITIALIZED BY RLUXGO FROM SEEDS    2    0    0

 NCALLS =  100000000

 XSECT FOR WEIGHTED EVENTS

 EFF =  0.999998000004
 XSECT =  13.52642598410257 +-  2.200641628422248E-03 (PB)

 ENERGY LOSS
 MEAN EG =  1.18064898022852 +-  6.041171111535717E-04 (GEV)

 INVARIANT-MASS LOSS
 MEAN IML =  1.17777558571609 +-  6.023399436481718E-04 (GEV)

\end{verbatim}

\vfil\eject
\pagestyle{empty}

\leftline{\bf Sample 2}
\vskip 12pt

\begin{verbatim}

 SQRT(S) =  176.0  GEV
 M_Z =  91.1888  GEV
 G_Z =  2.4974  GEV
 M_W =  80.23  GEV
 G_W =  2.03367033063195  GEV
 STH2 =  0.2310309124510679

 GVE =  -1.409737267299689E-02
 GAE =  -0.185794027211796
 GWF =  0.230409927395451
 GWWZ =  -0.571479454308384

 RANLUX INITIALIZED BY RLUXGO FROM SEEDS   2   0   0

 NHIT =  10000

 XSECT FOR UNWEIGHTED EVENTS

 EFF =  1.341451799624662E-02
 XSECT =  13.4145179962466  +-  0.133242395670936 (PB)
 NBIAS/NMAX =  0.0

\end{verbatim}

\end{document}